\documentclass[twocolumn,showpacs,preprintnumbers,amsmath,amssymb]{revtex4}

\usepackage{graphicx}
\usepackage{dcolumn}
\usepackage{bm}
\usepackage{amssymb,wasysym}

\begin{document}


\title{4D Topological Mass by Gauging Spin}

\author{Ishita D. Choudhury}
\email{ishitadutta.choudhury@bose.res.in}
\affiliation{%
{S. N. Bose National Centre for Basic Sciences, Block JD, Sector III, Salt Lake, Kolkata 700098, India}
}%

\author{M. Cristina Diamantini}
\email{cristina.diamantini@pg.infn.it}
\affiliation{%
NiPS Laboratory, INFN and Dipartimento di Fisica, University of Perugia, via A. Pascoli, I-06100 Perugia, Italy
}%

\author{Giuseppe Guarnaccia}
\email{guarnacciagiuseppe@yahoo.it}
\affiliation{%
{Dipartimento di Fisica ``E.~R. Caianiello'', Universit\`a di
Salerno, I-84084 Fisciano (Salerno), Italy}
}%

\author{Amitabha Lahiri}
\email{amitabha@bose.res.in}
\affiliation{%
{S. N. Bose National Centre for Basic Sciences, Block JD, Sector III, Salt Lake, Kolkata 700098, India}
}%

\author{Carlo A. Trugenberger}
\email{ca.trugenberger@bluewin.ch}
\affiliation{%
SwissScientific, chemin Diodati 10, CH-1223 Cologny, Switzerland
}%


\date{\today}

\begin{abstract}
We propose a spin gauge field theory in which the curl of a Dirac fermion current density plays the role of
the pseudovector charge density. In this field-theoretic model, spin interactions are mediated by a single scalar gauge boson in its antisymmetric tensor formulation. We show that these long range spin interactions induce a gauge invariant photon mass in the one-loop effective action. The fermion loop generates a coupling between photons and the spin gauge boson, which acquires thus charge. This coupling represents also an induced, gauge invariant, topological mass for the photons, leading to the Meissner effect. The one-loop effective equations of motion for the charged spin gauge boson are the London equations. We propose thus spin gauge interactions as an alternative, topological mechanism for superconductivity in which no spontaneous symmetry breaking is involved. 

\end{abstract}
\maketitle

Topology plays a major role in both field theory and condensed matter systems. It is by now well known that it is not necessary to spontaneously break the U(1) gauge symmetry to generate a photon mass: the famed topological Chern-Simons (CS) term in (2+1) dimensions \cite{jackiw} accomplishes this in a gauge invariant manner, albeit breaking the $P$ and $T$ symmetries. Since the CS term is the infrared-dominant term in the gauge field action it can describe new universality classes of topological matter \cite{wen1} when the dual field strength is used to represent a conserved matter current. 

An analogous mechanism can give photons a gauge invariant mass in (3+1) dimensions, and this time without breaking $P$ and $T$. This is achieved by another topological term, called the $BF$ term and coupling the dual field strength $\tilde F_{\mu \nu}$ to a second-rank antisymmetric pseudotensor $ B_{\mu \nu}$ \cite{bowick}. This is a topological mechanism for superconductivity, in which the field equations for the antisymmetric tensor represent the London equations. 

Due to a "vector" gauge symmetry, leaving the action invariant under transformations $B_{\mu \nu}$ $\to $ $B_{\mu \nu} + \partial_{\mu}\lambda_{\nu} - \partial_{\nu}\lambda_{\mu}$, the antisymmetric tensor field encodes actually a single scalar degree of freedom. Classically, this is dual to the phase of a $U(1)$ order parameter and the superconductivity mechanism is dual to the Higgs mechanism in the limit of an infinitely heavy Higgs boson: in the $BF$ mechanism, it is actually the scalar boson that "eats up" the original massless photon to become a massive photon. Quantum mechanically, however, the theories are deeply different, since one can prove \cite{bowick} that no spontaneous symmetry breaking is involved in the $BF$ mechanism, contrary to the Higgs mechanism. The $BF$ superconductivity mechanism is a topological mechanism for superconductivity, genuinely different from the Higgs mechanism. 

It is well known that the CS term in (2+1) dimensions is radiatively induced at one-loop level\cite{redlich} even if it is not present in the original Lagrangian. The generated CS coupling constant, in this case, is proportional to the sign of the fermion mass.

In a previous publication \cite{wrong}, three of us proposed that the topological BF term is induced at one loop also in (3+1) dimensions by coupling the antisymmetric pseudotensor gauge field $B_{\mu \nu}$ gauge-invariantly to the vorticity field of charged fermions but, unfortunately, the details of the computation are not entirely correct. In this paper we present the correct results and we show that, essentially, the mechanism that generates the topological photon mass in (3+1) dimensions is a gauge theory of spin. It is known that spin interactions, and particularly spin-orbit coupling, play an important role in the physics of topological insulators \cite{galitski}. It has also been shown that collective excitations, like phonons, can induce long range spin-spin interactions \cite{bennet}. Here, we propose a spin gauge interaction as an alternative, topological mechanism for superconductivity. Of course this spin gauge theory has to be understood as a low energy effecti!
 ve theory valid for energy scales well below an ultraviolet cutoff $\Lambda$. Correspondingly, and contrary to the (2+1)-dimensional case, the induced topological coupling is cutoff-dependent, being proportional to  $m \ln {\Lambda^2 \over m^2}$, with $m$ the fermion mass. 
Topological mass generation in (3+1)-dimensional QED (without the $B_{\mu\nu}$ tensor field) has been also studied in \cite{dvali} within a generalisation to 4-dimensions of the Schwinger model.

First attempts to extend the gauge principle to vector-like charges \cite{leblanc} have considered as obvious candidates for a tensor gauge theory of spin the standard Dirac tensor and pseudotensor densities $\bar \psi \sigma^{\mu \nu} \psi$ and $\bar \psi \gamma^5  \sigma^{\mu \nu} \psi$, with 
$\sigma^{\mu \nu} \equiv (i/2) [\gamma^{\mu} , \gamma^{\nu}]$ (we shall use units in which $\hbar =1$, $c=1$ throughout the paper, $ [ , ]$ denotes  commutators and  $ \{ , \}$  anticommutators). Both, however are not suitable for a gauge theory, since they are not conserved. 

The spin density of Dirac fermions is given by the expression $\psi^{\dagger} \bf \Sigma \psi$ with $\bf \Sigma = {\rm diag} ({\bf \sigma}, {\bf \sigma})$. Using $\sigma^{ij} = \epsilon^{ijk} \Sigma_k$, this can be embedded in the spin current $S^{\mu , \alpha \beta} = (1/4) \bar \psi \{ \gamma^{\mu}, \sigma^{\alpha \beta} \} \psi$. This, however, is a third-order tensor. In order to obtain a second-order tensor we shall consider the non-local field generated by the curl of the spin current. Moreover, in order to maintain $P$ and $T$ conservation when coupling to the pseudotensor $B_{\mu \nu}$, we shall add a $i\gamma^5$ matrix in the original spin current. We shall thus couple $B_{\mu \nu}$ as 
\begin{equation}
{\cal L}^B_{\rm int} = g B_{\mu \nu} J^{\mu \nu}
\label{zero}
\end{equation}
to the pseudotensor current 
\begin{equation}
J^{\mu \nu} = { -m\over 2} {\partial_{\alpha}\over \Box  }\left( \bar \psi \gamma^5 \{ \gamma^{\alpha}, \sigma^{\mu \nu} \} \psi\right) =
{- im\over 6}  {\partial_{\alpha }\over \Box} \left( \bar \psi \gamma^5 \gamma^{ [  \alpha } \gamma^{\mu} \gamma^{\nu ] } \psi \right)\ ,
\label{one}
\end{equation}
where the symbol $[ \dots ]$ in the exponent of [\ref{one}] denotes total antisymmetrization of the indices and $g$ is a dimensionless coupling. 
In the second representation of this current, its conservation is explicit. Thus, the coupling $B_{\mu \nu} J^{\mu \nu}$ is invariant under "vector" gauge transformations 
\begin{equation}
B_{\mu \nu} \to B_{\mu \nu} + \partial_{\mu} \lambda_{\nu} - \partial_{\nu} \lambda_{\mu} \ .
\label{two}
\end{equation}
It is also invariant under $P$ and $T$ transformations. 

In order to establish the physical nature of this coupling let us use the identity
\begin{equation}
\gamma^{ [  \alpha } \gamma^{\mu} \gamma^{\nu ] } =  6 i \ \epsilon^{\alpha\mu\nu\sigma} \gamma^5 \gamma_\sigma \ ,
\label{twoaddone}
\end{equation}
to rewrite the current $J^{\mu \nu}$ as
\begin{equation}
J^{\mu \nu} = m \ \epsilon^{\alpha\mu\nu\sigma} {\partial_{\alpha }\over \Box} J_{\sigma} \ , 
\label{twoaddtwo}
\end{equation}
where $J^{\mu} = \bar \psi \gamma^{\mu} \psi$ is the usual Dirac fermion current. The components $J^{0i}$ that couple to $B_{0i}$ are thus given by $J^{0i}= -m\ \epsilon^{ijk} (\partial_j / \Box) \psi^{\dagger} \alpha^k \psi$. The expression $\psi^{\dagger} \alpha^k \psi$ is the velocity field of the Dirac fermion. Thus, in analogy to fluid dynamics, the pseudovector charge density $\Box J^{0i}$, given by the curl of the velocity field, represents the vorticity field of the fermion. 

Using the Gordon decomposition (for the purpose of illustrating the spin dependence of the current, we consider the Gordon decomposition of the  non-interacting case)
\begin{equation}
\psi^{\dagger} \alpha^k \psi = {i\over 2m} \left[ \bar \psi \partial^k \psi - \partial^k \bar \psi \psi \right] 
+{1\over 2m} \partial_{\mu} \ \bar \psi \sigma^{k \mu} \psi \ ,
\label{three}
\end{equation}
one can separate the pseudovector charges into their orbital and spin contributions. In the non-relativistic limit, in which the lower components of Dirac spinors can be neglected for energies much lower than their mass, this reduces to \cite{greiner} 
\begin{equation}
\left( \psi^{\dagger} \alpha^k \psi \right)_{\rm NR} \to {i\over 2m} \left[ \bar \phi \partial^k \phi - \partial^k \bar \phi \phi \right] + {1\over 2m} \epsilon^{kij} \partial_i  \ \phi^{\dagger} \sigma^j \phi \ ,
\label{four}
\end{equation}
where the spinors $\phi$ on the right-hand side are the two-dimensional Pauli spinor corresponding to the upper components of the four-dimensional Dirac spinors $\psi$. In the static case, in which pseudovector charges are time-independent and cannot accumulate (in analogy to the usual magnetostatic case of scalar charges), 
the non-relativistic spin contribution to the pseudovector charge density becomes thus
\begin{equation}
\left( J^{0i}_{\rm spin} \right)_{\rm NR} = {1\over 2}  \left( \psi^{\dagger} \sigma^i \psi \right) \ ,
\label{five}
\end{equation}
which is the intrinsic magnetic moment density of the particle. This is why we speak of gauging the spin. 

We shall henceforth consider a fully relativistic vortex gauge model of massive (mass $m$) charged (charge $-e$) Dirac fermions with the following Lagrangian density
\begin{eqnarray}
{\cal L} &&= \bar \psi \gamma^{\mu} \left(i \partial_{\mu} + eA_{\mu} \right) \psi - m \bar \psi \psi
+  g B_{\mu \nu} J^{\mu\nu} 
\nonumber \\
&&-{1\over 4 } F_{\mu \nu}F^{\mu \nu} + {1\over 12} H_{\mu \nu \alpha} H^{\mu \nu \alpha} \ ,
\label{six}
\end{eqnarray} 
where $H_{\mu \nu \alpha} \equiv \partial_{[ \mu} B_{\nu \alpha ] } $ is the Kalb-Ramond \cite{kalb} gauge invariant field strength for the antisymmetric tensor and $g$ is the dimensionless vortex coupling constant. As anticipated, this model has to be considered as a low-energy effective theory, valid for energy scales below an ultraviolet cutoff $\Lambda$. Note that the antisymmetric tensor interaction can be easily rendered local by means of Lagrange multipliers $\lambda_{\mu \nu}$,
\begin{equation}
{\cal L}^B_{\rm int} = gB_{\mu \nu} \tilde J^{\mu \nu} +\lambda_{\mu \nu} \left( { -m\over 2} \partial_{\alpha} \left( \bar \psi \gamma^5 \{ \gamma^{\alpha}, \sigma^{\mu \nu} \} \psi\right) 
-\Box \tilde J^{\mu \nu}\right) \ .
\label{sixadd}
\end{equation}
It is this local formulation that must be adopted to count degrees of freedom. Let us do so for the gauge degrees of freedom. As usual, the massless vector gauge field $A_{\mu}$ embodies two degrees of freedom. 
Due to the structure of the Kalb Ramond kinetic term, the mixed components $B_{0i}=-B_{i0}$ play the role of non-dynamical Lagrange multipliers, leaving three dynamical degrees of freedom in the tensor $B_{\mu \nu}$. The gauge invariance eq. (\ref{two}) eliminates, however, two of these, since there are two independent gauge parameters $\lambda_i$ (the other one being eliminated by the equivalence $\lambda_i \equiv \lambda_i + \partial _i \eta $), leaving thus one overall degree of freedom. The long-range vortex gauge interaction is thus mediated by a single massless scalar boson. In addition to this, however, there are two massless degrees of freedom, one in the conserved current $\tilde J^{\mu \nu}$ and one in the gauge-invariant Lagrange multiplier $\lambda_{\mu \nu}$. These can be considered as auxiliary degrees of freedom needed for a local formulation of spin gauge interactions. 

For calculations, however, we shall stick with the non-local formulation and rewrite the antisymmetric tensor interaction as
\begin{equation} 
{\cal L}^B_{\rm int} = g B_{\mu \nu} J^{\mu\nu}  = {2mg \over \Box} F_\mu J^\mu \ ,
\label{fua}
\end{equation}
where $F_\mu = (1/2) \epsilon_{\mu\nu\alpha\beta} \partial^\nu B^{\alpha\beta}$ is the dual Kalb-Ramond field strength. Formally, thus, the inverse d'Alembertian of the dual Kalb-Ramond field strength plays the role of a gauge field in the Lorenz gauge and we can rewrite the total interaction term of the Fermions as
\begin{eqnarray}
{\cal L}_{\rm int} &&= e (A_{\rm eff})_{\mu} J^{\mu} \ , 
\nonumber \\
 (A_{\rm eff})_{\mu} &&=  A_{\mu} + {2mg\over e\Box} F_{\mu}  \ .
\label{totint}
\end{eqnarray}
In order to integrate out the fermions and obtain the one-loop induced action we need now only the standard vacuum polarisation tensor. In momentum space this is given by 
\begin{eqnarray}
& \Gamma^{(\rm 1-loop)} = - \frac{ie^2}{2}Tr(\frac{1}{i{\not}\partial-m}{\not} A_{\rm eff}\frac{1}{i{\not}\partial-m}{\not} A_{\rm eff}) \nonumber \\
&= \frac{1}{2}\varint\frac{d^4k}{(2\pi)^4}(A_\mu)_{\rm eff} (-k)(A_\nu)_{\rm eff}(k)\Pi^{\mu\nu}(k)
\label{AA3}
\end{eqnarray}
where $\Pi^{\mu\nu}$ is the usual QED vacuum polarization tensor 
\begin{eqnarray}
&\Pi^{\mu\nu}(k)=(g^{\mu\nu}k^2-k^\mu k^\nu)\Pi(k^2)\nonumber \\
& \Pi(k^2)  = \frac{e^2}{12\pi^2}\ln\frac{\Lambda^2}{m^2} \ ,
\label{aa2}
\end{eqnarray}  
and $\Lambda$ is the momentum space ultraviolet cutoff. 

We obtain thus the 1-loop induced action
\begin{equation}
\Gamma^{\rm 1-loop} =  \frac{e^2}{4}\frac{1}{12\pi^2}\ln\frac{\Lambda^2}{m^2} \varint d^4x(F_{\rm eff})_{\mu\nu}(F_{\rm eff})^{\mu\nu} \ .
\label{AA}
\end{equation}
with 
\begin{equation}
(F_{\rm eff})_{\mu \nu} = F_{\mu \nu} + {2mg\over e} \left( {\partial_{\mu} \over \Box} F_{\nu}-{\partial_{\nu} \over \Box} F_{\mu} \right) \ .
\label{effF}
\end{equation}
Expanding this expressions leads to three terms in the original variables
\begin{equation}
\Gamma^{\rm 1-loop} = \Gamma^{\rm 1-loop}_{AA} + \Gamma^{\rm 1-loop}_{AB}+\Gamma^{\rm 1-loop}_{BB}\ ,
\label{exp}
\end{equation}
where
\begin{eqnarray}
\Gamma^{\rm 1-loop}_{AA} &&=  \frac{e^2}{4}\frac{1}{12\pi^2}\ln\frac{\Lambda^2}{m^2} \varint d^4x \ F_{\mu\nu}F^{\mu\nu} \ ,
\label{indAA} \\
\Gamma^{\rm 1-loop}_{AB} &&=  \frac{8mg}{4}\frac{1}{12\pi^2}\ln\frac{\Lambda^2}{m^2} \varint d^4x \ A_{\mu}F^{\mu} \ ,
\label{indAB} \\
\Gamma^{\rm 1-loop}_{BB} &&=  \frac{8m^2g^2}{4}\frac{1}{12\pi^2}\ln\frac{\Lambda^2}{m^2} \varint d^4x \ F_{\mu} {1\over \Box} F^{\mu} \ .
\label{indBB}
\end{eqnarray}
We can then combine this induced action with the original one to obtain the 1-loop effective Lagrangian. With the usual rescaling $A_\mu \rightarrow e A_\mu$, to bring the action in its standard form, we obtain,
\begin{eqnarray}
{\cal L}  &&=  -{1\over 4 e_{\rm ph}^2 } F_{\mu \nu}F^{\mu \nu} + {g_{\rm ph}m \over 2 \pi} B_{\mu\nu} \epsilon^{\mu\nu\rho\sigma}\partial_\rho A_\sigma + {1\over 12} H_{\mu \nu \alpha} H^{\mu \nu \alpha}  
\nonumber \\
&&+ {6g_{\rm ph}^2m^2\over \ln {\Lambda^2 \over m^2}} F_{\mu} {1\over \Box} F^{\mu} \ ,
\label{a1l}
\end{eqnarray}
where
\begin{equation}
e^2_{\rm ph} = e^2 \left( 1 + {e^2 \over 12 \pi^2} \ln {\Lambda^2 \over m^2} \right) \ ,
\label{eph}
\end{equation}
is the QED renormalized charge and 
\begin{equation}
g_{\rm ph} = {g\over 6 \pi}  \ln {\Lambda^2 \over m^2}  \ .
\label{msc}
\end{equation}
Finallly, we can represent the induced non-local term as a Gaussian integral over a new massless vector gauge field $C_{\mu}$ (a regulator mass is needed which can be eliminated at the end). This gives
\begin{eqnarray}
{\cal L}  &&=  -{1\over 4 e_{\rm ph}^2 } F_{\mu \nu}F^{\mu \nu} + {g_{\rm ph} m\over 2 \pi} B_{\mu\nu} \epsilon^{\mu\nu\rho\sigma}\partial_\rho A_\sigma + {1\over 12} H_{\mu \nu \alpha} H^{\mu \nu \alpha}  
\nonumber \\
&& - {\ln {\Lambda^2 \over m^2}\over 48 \pi^2} G_{\mu\nu} G^{\mu \nu} +{g_{\rm ph}m \over 2 \pi} B_{\mu\nu} \epsilon^{\mu\nu\rho\sigma}\partial_\rho C_\sigma \ ,
\label{local}
\end{eqnarray}
where $G_{\mu \nu}$ is the field strength corresponding to the new gauge field $C_{\mu}$.
At this point we can diagonalize the action by introducing the gauge field combinations
\begin{eqnarray}
A_{\mu} && \to  A_{\mu} + C_{\mu} \ ,
\nonumber \\
C_{\mu } &&\to  A_{\mu } -{4e_{\rm ph}^2 \over 12 \pi^2}  \ln {\Lambda^2 \over m^2} \ C_{\mu} \ ,
\label{diag}
\end{eqnarray}
The resulting 1-loop Lagrangian decouples into two sectors, 
\begin{eqnarray}
{\cal L} &&= {\cal L}_1 + {\cal L}_2 \ ,
\nonumber \\
{\cal L}_1  &&=  -{1\over 4 e_1^2 } F_{\mu \nu}F^{\mu \nu} + {g_{\rm ph} m\over 2 \pi} B_{\mu\nu} \epsilon^{\mu\nu\rho\sigma}\partial_\rho A_\sigma + {1\over 12} H_{\mu \nu \alpha} H^{\mu \nu \alpha}  \ ,
\nonumber \\
{\cal L}_2 &&=- {1\over 4e_2^2} G_{\mu\nu} G^{\mu \nu}  \ ,
\label{decom}
\end{eqnarray}
where 
\begin{eqnarray}
e_1^2 &&= e_{\rm ph}^2 {1+{4e_{\rm ph}^2 \over 12 \pi^2}\ln {\Lambda^2 \over m^2} \over {4e_{\rm ph}^2 \over 12 \pi^2}\ln {\Lambda^2 \over m^2}} \ ,
\nonumber \\
e_2^2 &&= e_{\rm ph}^2 \left( 1+{4e_{\rm ph}^2 \over 12 \pi^2}\ln {\Lambda^2 \over m^2} \right) \ .
\label{newcou}
\end{eqnarray}
The first sector is a topological BF model with an effective charge $e_1$ that reduces to the usual renormalized electron charge, $e_1 \to e_{\rm ph}$, for $m \ll \Lambda$. Contrary to the (2+1)-dimensional case, the topological coupling constant $g_{\rm ph}$ is also cutoff-dependent via eq. (\ref{msc}). The second sector contains two decoupled massless degrees of freedom that correspond to the two auxiliary degrees of freedom in the additional current and the Lagrange multiplier of the local formulation of the fermion interaction, eq. (\ref{sixadd}). 

The equations of motions for the topological sector are
\begin{eqnarray}
&\partial_\mu F^{\mu\nu} = - {e_1^2  g_{\rm ph}m \over 6 \pi} \epsilon^{\nu\mu\alpha\beta} H_{\mu\alpha\beta} \ , \nonumber \\
&\partial_\mu H^{\mu\alpha\beta}  = {g_{\rm ph} m\over 2 \pi} \epsilon^{\alpha\beta\mu\nu} F_{\mu\nu} \ ,
\label{var}
\end{eqnarray}
from which one can derive  \cite{bowick,diam}:
\begin{equation}
\left[ \Box + \left( { e_1 g_{\rm ph} m \over \pi} \right)^2 \right] F_{\mu\nu} = 0 \ .
\label{efm}
\end{equation}
Eq.(\ref{efm}) shows that the field strength $F_{\mu\nu} $ satisfies the Klein-Gordon equation with mass given by ${ e_1  g_{\rm ph} M \over \pi} $. By the same analysis also $H^{\mu \nu \alpha} $ satisfies the same equation of motion.

>From eq.(\ref{var}) we also recognize that the dual of the Kalb-Ramond field strength acts as charged current for the photon field:
\begin{equation}
\partial_\mu F^{\mu\nu} =  J^\nu\ ,   J^\nu \equiv  {- e_1^2  g_{\rm ph}m \over  \pi}H^\nu =  {- e_1^2  g_{\rm ph}m\over 6 \pi} \epsilon^{\nu\mu\alpha\beta} H_{\mu\alpha\beta}  \ .
\label{var1}
\end{equation}
Substituing eq.(\ref{var1}) in the second equation  eq.(\ref{var}) we obtain:
\begin{eqnarray} 
&\epsilon^{\mu\nu\alpha\beta} \partial_\alpha J_\beta =  - \left( {e_1 g_{\rm ph} m\over  \pi} \right)^2  \tilde F_{\mu\nu}  \ , \nonumber \\
&\tilde F_{\mu\nu} = {1\over 2} \epsilon^{\mu\nu\alpha\beta} F_{\alpha\beta}
\label{var2}
\end{eqnarray}
which is nothing else than a relativistic version of the London equations for superconductivity.   

In conclusion, we have shown that a relativistic model of spin gauge interactions mediated by a scalar boson in an effective low-energy field theory generates the BF topological mass term at 1-loop in the effective action. 
Photons acquire a gauge-invariant topological mass, the additional scalar boson acquires charge and the corresponding current satisfies the London equations, all the hallmarks of "gauge-invariant" superconductivity  with no spontaneous symmetry breaking involved.

\end{document}